\documentclass[5p]{elsarticle}

\usepackage{hyperref}
\usepackage{graphicx}

\journal{Nuclear Instruments and Methods in Physics Research Section B\ }

\begin{document}

\begin{frontmatter}

\title{Response of a proportional counter to $^{37}$Ar and $^{71}$Ge:\\
Measured spectra versus Geant4 simulation}

\author[inr]{D.N.~Abdurashitov\corref{cor}}
\ead{jna@inr.ru}
\author[inr,cuc]{Yu.M~Malyshkin}
\author[inr]{V.L.~Matushko}
\author[pri]{B.~Suerfu}

\address[inr]{Institute for Nuclear Research, prospekt 60-letiya Oktyabrya 7a, Moscow 117312, Russia}
\address[cuc]{Pontifical Catholic University of Chile, Santiago, Chile}
\address[pri]{Princeton University, NJ 08544, USA}

\cortext[cor]{Corresponding author}

\begin{abstract}
The energy deposition spectra of $^{37}$Ar and $^{71}$Ge in a miniature proportional counter are measured and compared in detail to the model response simulated with Geant4. A certain modification of the Geant4 code, making it possible to trace the deexcitation of atomic shells properly, is suggested. Modified Geant4 is able to reproduce a response of particle detectors in detail in the keV energy range. This feature is very important for the laboratory experiments that search for massive sterile neutrinos as well as for dark matter searches that employ direct detection of recoil nuclei. This work demonstrates the reliability of Geant4 simulation at low energies.
\end{abstract}


\end{frontmatter}


\section{\label{sec:intro}Introduction}

Nonzero neutrino mass and the existence of dark matter, whose nature is not known yet, indicate that the Standard Model of particles is incomplete. A sterile neutrino is one of the most natural candidates for dark matter particles~\cite{abaz_WP:2012}. From this point of view direct laboratory searches for sterile neutrinos are of a particular interest. An experiment aiming to search for a hypothetical admixture of sterile neutrinos in $\beta$--decay of tritium was proposed~\cite{jna:2015r1}. The admixture can be detected as a specific distortion of the tritium electrons energy spectrum in a proportional counter. The distortion of interest is expected to be very small, therefore it is very important to investigate every possible source of a systematic uncertainty. To do this, we have simulated the response of a proportional counter to the decay of $^{37}$Ar and $^{71}$Ge and compared it to the measured spectra. We found that the current version of Geant4~\cite{geant:2003} (10.1 patch 2) used for simulation does not reproduce all specific features of the energy deposition spectra from those sources. We found also that it was because Geant4 did not trace accurately a fundamental phenomenon namely the filling of atomic shell vacancies.  We suggested a code modification (patch) that describes the deexcitation of atomic shells correctly. In this work we present the measured energy spectra in comparison with the model responses before and after applying the patch.

\section{\label{sec:experiment}Measured spectra}
\subsection{\label{subsec:scheme}Decay of $^{37}$Ar and $^{71}$Ge}


\begin{table*}%
\begin{center}
\caption{\label{tab:scheme}Radiation produced in the decay of $^{37}$Ar (left) and $^{71}$Ge (right). The energies of Auger electrons and X-rays are expressed in keV.}%
\begin{small}
\begin{tabular*}{\textwidth}{c@{\extracolsep{\fill}}cccccc}
\hline\hline
 Decay & Sum energy of   & Energy of & Percent of  & Sum energy of   & Energy of & Percent of\\
  mode & Auger electrons &   X-ray   & all decays  & Auger electrons &   X-ray   & all decays\\
\cline{1-1}\cline{2-4}\cline{5-7}
  K    & 2.823  & 0.0   & 81.5 & 10.367 & 0.0    & 41.4\\
  K    & 0.202  & 2.621 & 2.7  & 1.143  & 9.224  & 13.7\\
  K    & 0.201  & 2.622 & 5.5  & 1.116  & 9.251  & 27.4\\
  K    & 0.007  & 2.816 & 0.5  & 0.107  & 10.260 & 1.7\\
  K    &        &       &      & 0.103  & 10.264 & 3.5\\
\cline{1-1}\cline{2-4}\cline{5-7}
  L    & 0.270  & 0.0 & 8.9    & 1.299  & 0.0    & 10.3\\
\cline{1-1}\cline{2-4}\cline{5-7}
  M    & 0.018  & 0.0 & 0.9    & 0.160  & 0.0    & 2.0\\
\hline\hline
\end{tabular*}
\end{small} 
\end{center}
\end{table*}

To understand the methods of measurement and simulation, it is necessary to know the radiations that are produced after the decay of the two isotopes that were employed. $^{37}$Ar and $^{71}$Ge atoms decay by electron capture followed by deexcitation of the daughters $^{37}$Cl and $^{71}$Ga. Note that in a gaseous proportional counter only the total energy of all Auger electrons is detected with no distinguishing of separate Auger-lines.

$^{37}$Ar decays by K-, L-, and M-electron capture.  The ratios of these modes are L/K = 0.0987 \cite{huster:1969} and M/L = 0.104 \cite{renier:1968}.  This leads to absolute fractions K = 0.9017, L = 0.0890, and M = 0.0093. Taking into account the X-ray branching ratios and the X-ray energies arising from vacancies in the various shells (see, for example, \cite{Lederer:1978}) leads to the left part of the Table~\ref{tab:scheme}. If the X-rays and Auger electrons deposit all their energy in the counter, three main peaks at 2.82, 0.27 and 0.02~keV will appear in a pulse height spectrum. Those peaks, named  as K-, L- and M-peak according to their origin, are used to calibrate an energy scale in measurements.

In case of $^{71}$Ge decay the absolute fractions of K-, L- and M-captures are about 0.880, 0.103 and 0.017, respectively \cite{renier:1971}. Further, taking into account the X-ray branching ratios and the X-ray energies leads to the right part of the Table~\ref{tab:scheme}. Again, if the X-rays and Auger electrons are all captured  in the counter, three peaks at 10.37, 1.2 and 0.16~keV will appear there. Note that the ``L-peak'' is not really a single peak but is a structure that contains three unresolved lines at 1.299, 1.143, and 1.116 keV. The height of the two last-mentioned components will vary with the probability for capture of the associated X-rays.  Thus the width of the L-peak, and to some extent its average energy, depend on the gas composition and the counter dimensions. If none of the X-rays are captured, the weighted average energy of the L-peak is 1.16 keV.

\subsection{\label{subsec:real}Detector and measurements}

A miniature proportional counter made from a fused silica tube, as described in \cite{kuzminov:1990} and \cite{yants:1994}, was used for the measurements. The tube is a cylinder with length 40~mm, inner diameter 4~mm, and wall thickness 1~mm. The inner surface of the tube is covered with 1~$\mu$m layer of pyrolithic carbon that serves as a cathode. A 12~$\mu$m tungsten anode wire is stretched along the cylinder axis. Similar counters are used in the SAGE solar neutrino experiment \cite{jna:sage2009} and demonstrate a high sensitivity to X-ray and Auger electron emission together with a high long-term stability.

The energy deposition spectrum after the $^{71}$Ge decay has been acquired from the counter filled with a mixture of Xe+11.0\% GeH$_4$ at the pressure of 703~Torr and the avalanche gain of about 10$^3$. The intensity of the $^{71}$Ge portion at the beginning of measurement was about 300~Bq. The measurement was done at the SAGE deep underground site in a passive shield \cite{jna:sage2009}. The background of the counter in this condition was less than 10 counts per day above 0.5~keV energy range. The $^{37}$Ar spectrum was measured in similar counter filled with pure Xe doped with tiny amount of $^{37}$Ar at the pressure of 760~Torr and with an avalanche gain of about 5$\cdot$10$^3$; the intensity was 540~Bq. The proportional counter anode was directly connected to a charge-sensitive preamplifier with the rise-time of 3~ns and decay-time of 30~$\mu$s. The shaping time of the readout electronics was about 1~$\mu$s for both measurements.

\begin{figure*}
\includegraphics[width=0.46\textwidth]{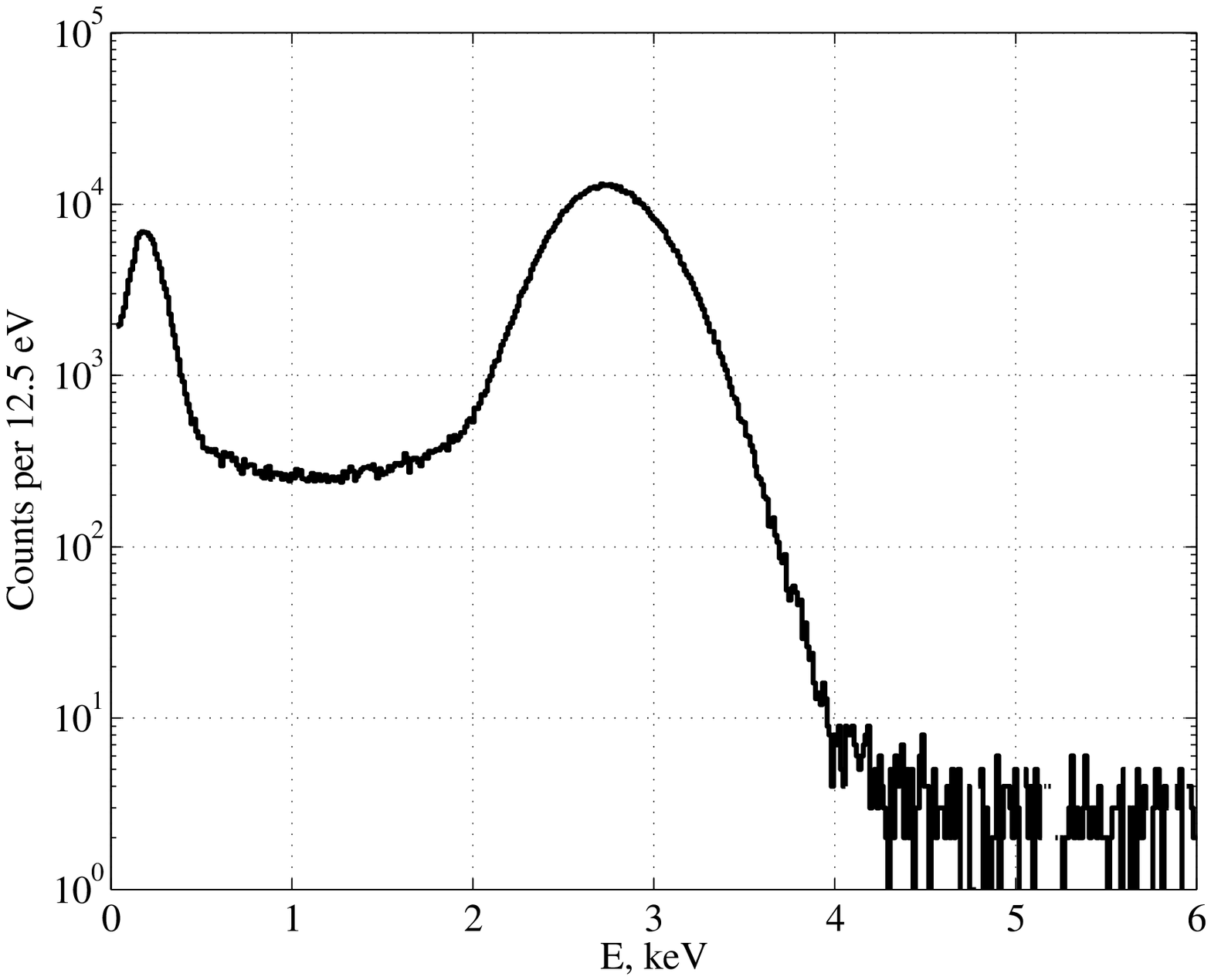}
\hfill
\includegraphics[width=0.46\textwidth]{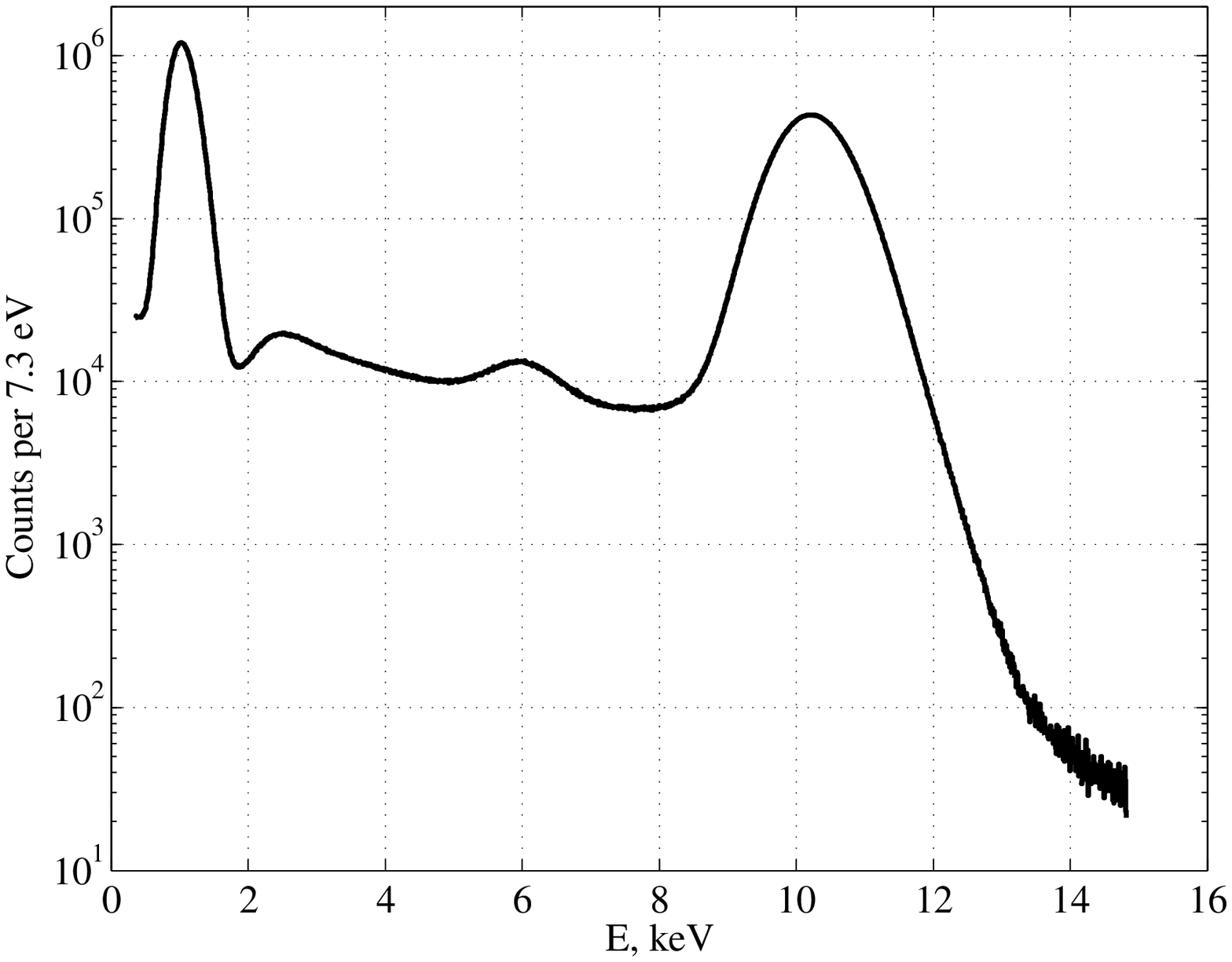}
\caption{\label{fig:meas} The energy deposition spectra of $^{37}$Ar (left) and $^{71}$Ge (right) measured in the proportional counter.}%
\end{figure*}

Collected pulse height spectra are shown in Fig.~\ref{fig:meas}. The total number of events is about 10$^6$ for $^{37}$Ar and about 1.7$\cdot$10$^8$ for $^{71}$Ge. The threshold in the measurements corresponded to 0.1~keV for $^{37}$Ar and 0.2~keV for $^{71}$Ge, so the M-peak remains invisible for both isotopes.

\section{\label{sec:simul}Simulation technique}

\subsection{\label{subsec:geant}Monte Carlo modeling}

The modeling was performed with the Geant4 software~ \cite{geant:2003}. Geant4 is a Monte Carlo code  widely used for simulation of particles' interaction and transport in high energy physics, medical physics, space research, and many other fields. Geant4 is an open source project, which allows users to easily adopt it for their own needs, e.g. implementing new particles and interactions. If the new features are expected to be useful for other users, they can be included in the subsequent versions of Geant4.

The geometry of the real proportional counter described above was implemented in a Geant4-based code. The counter was filled with a gas corresponding to each measurement, and the $^{37}$Ar or $^{71}$Ge sources were uniformly distributed throughout the counter volume. The low energy interactions were modeled with the Penelope model set~\cite{penelope:2001}, which uses EADL (Evaluated Atomic Data Library), EEDL (Evaluated Electronics Data Library) and EPDL97 (Evaluated Photon Data Library). The following processes were modeled: photon interactions (Compton effect, Rayleigh scattering, photo-effect and electron-positron pair production), interactions of electrons and positrons (bremsstrahlung and ionization) and atomic deexcitation. The Penelope developers claim the model is applicable for energies from 250~eV and higher, for all the elements $Z=1-100$, and the atomic deexcitation works only for $Z>5$, which is caused by limitations of EADL. Therefore, we didn't take into account the electric field because it is practically very hard to simulate low energy electron avalanches taking place in the vicinity of the anode. However, the main effect of the electric field presence, the so called ``end effect'', can be easily taken into account by another way, as discussed below. 

Results of Monte Carlo modeling are written to ROOT-files~\cite{root:1997}, where the energy deposition inside the counter and other values are collected for each decay. One billion decays were simulated for each source.

\subsection{\label{subsec:syst}Systematic distortion of model response}

A correct comparison of the measured pulse height spectrum to the model response is only possible after modeling systematic distortion of the simulated one. The sources of systematic distortions for a proportional counter are well known. The main systematic factor is a broadening of pulse heights caused by the statistical dispersion of the number of primary ion-electron pairs. Energy resolution $R$ is defined as the full width at half maximum (FWHM) of the response divided by the energy $E$: $R = {\rm FWHM}/{E}$. In case of normal distribution its dispersion $\sigma$ and FWHM are related as FWHM=2.34$\sigma$. The analytic expression of the response distortion can be written as
\[
S_{\scriptscriptstyle R}(E)=\int_0^{\infty} S(x) G(\mu=0, \sigma(x), x-E) dx,
\label{eq:resolution}   
\]
where $S(E)$ is the original normalized electron energy spectrum, $G(\mu,\sigma,E)$ is the normal (gaussian) distribution with $\sigma$ depending on $E$, and $S_{\scriptscriptstyle R}(E)$ is the final distorted response as a result of convolution. In this particular measurement $R$ was about 25\% and 13\% for the energies of 2.8 and 10.4~keV, respectively.

%
%
\begin{figure*}[!t]
\includegraphics[width=0.46\textwidth]{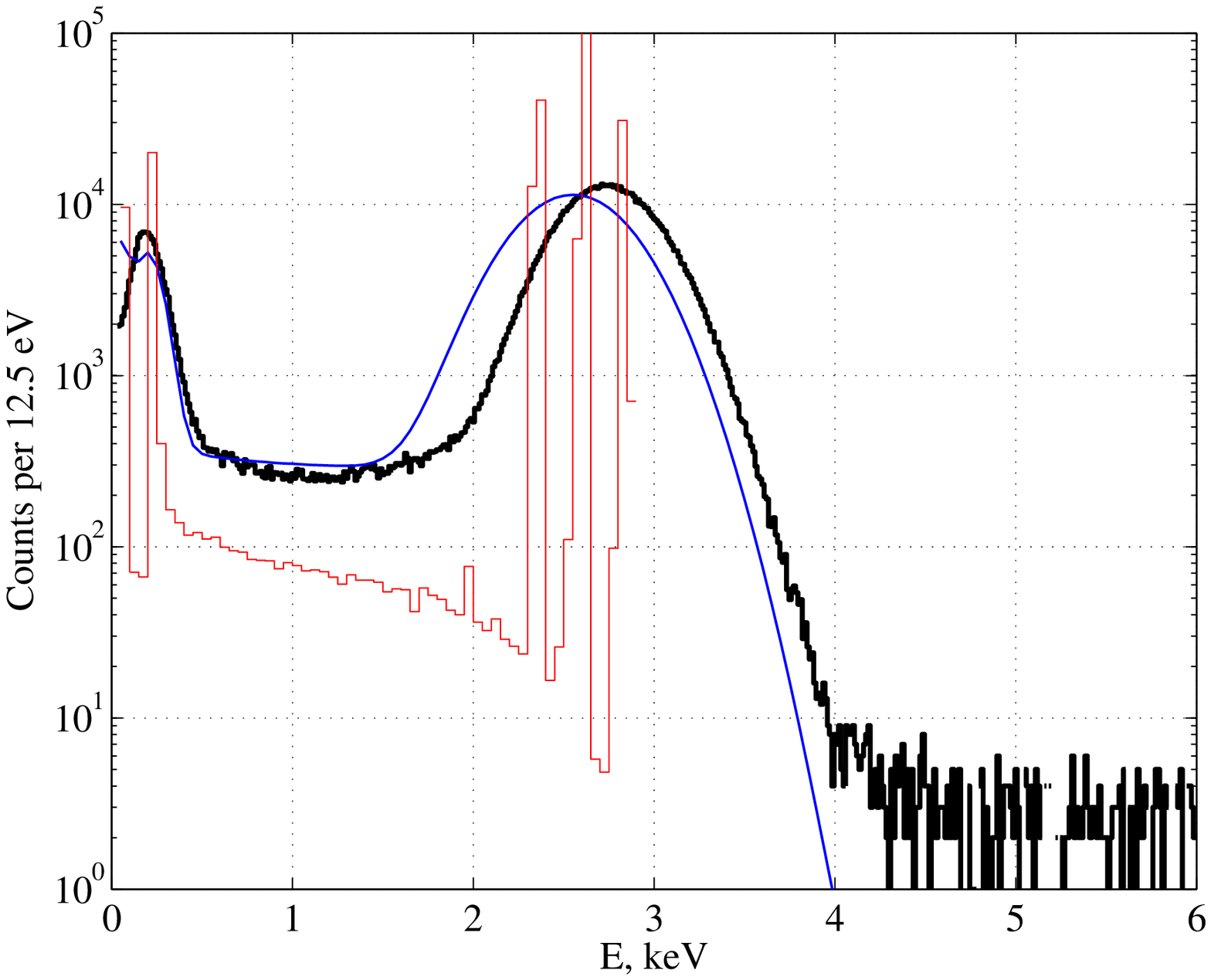}
\hfill
\includegraphics[width=0.46\textwidth]{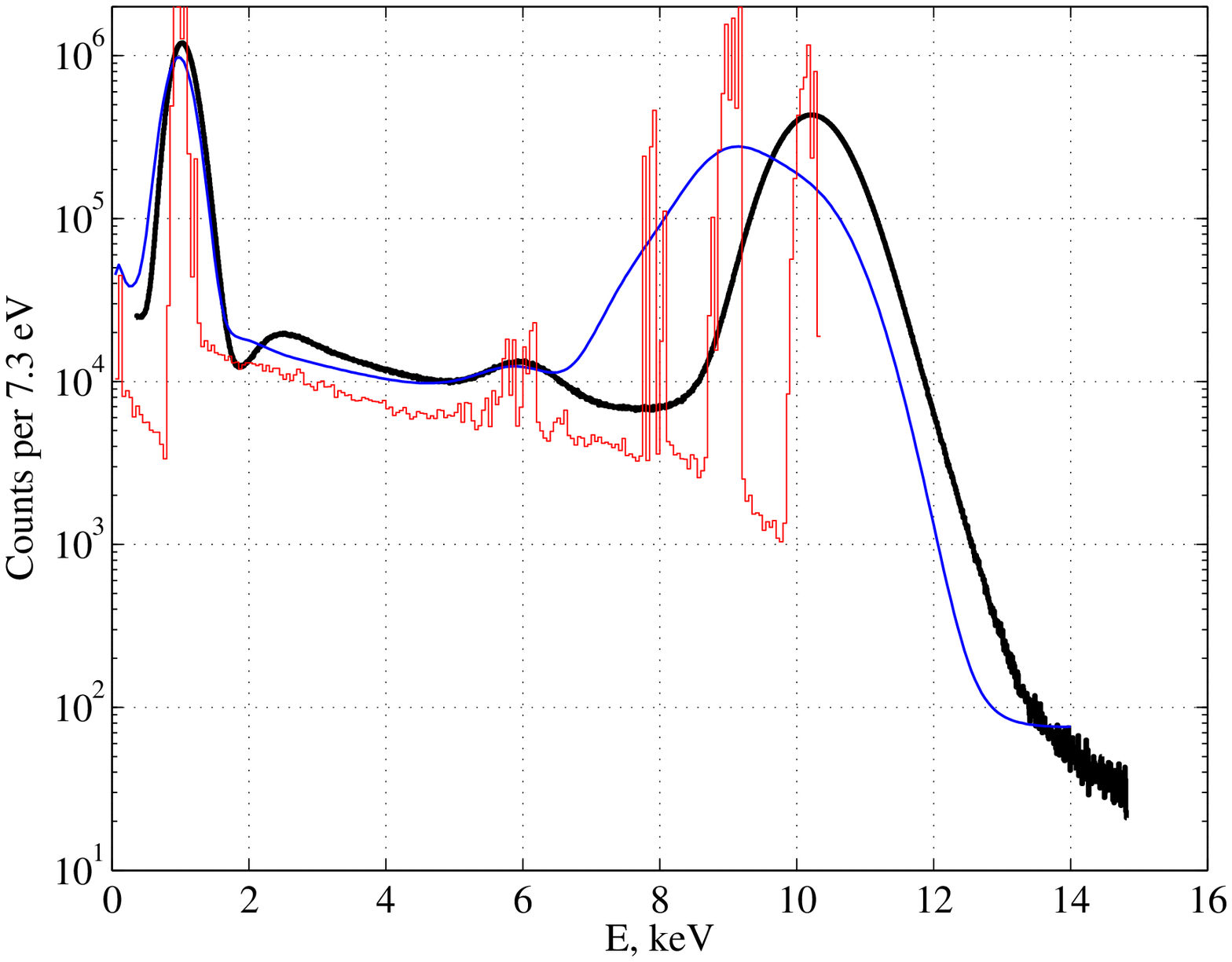}
\includegraphics[width=0.46\textwidth]{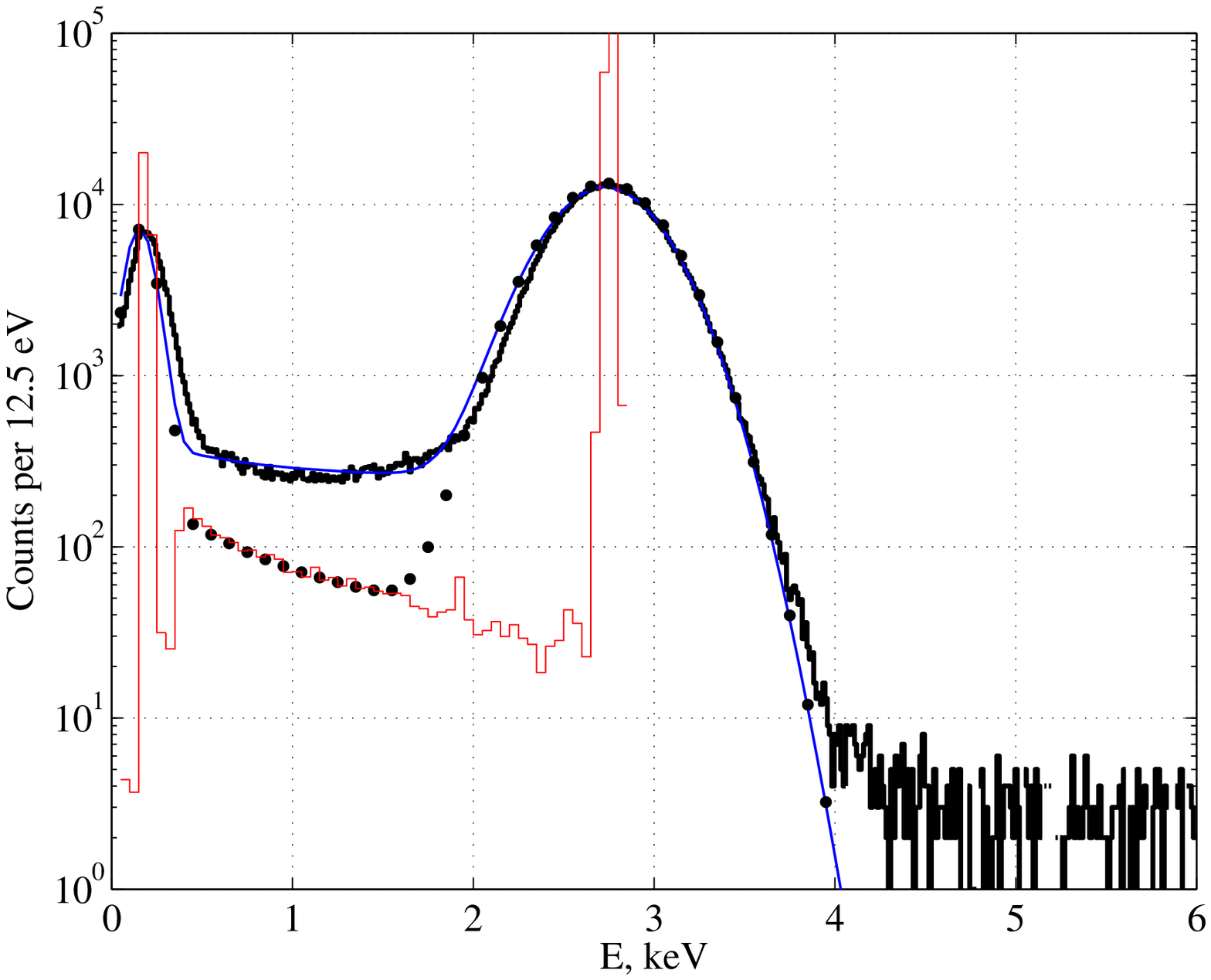}
\hfill
\includegraphics[width=0.46\textwidth]{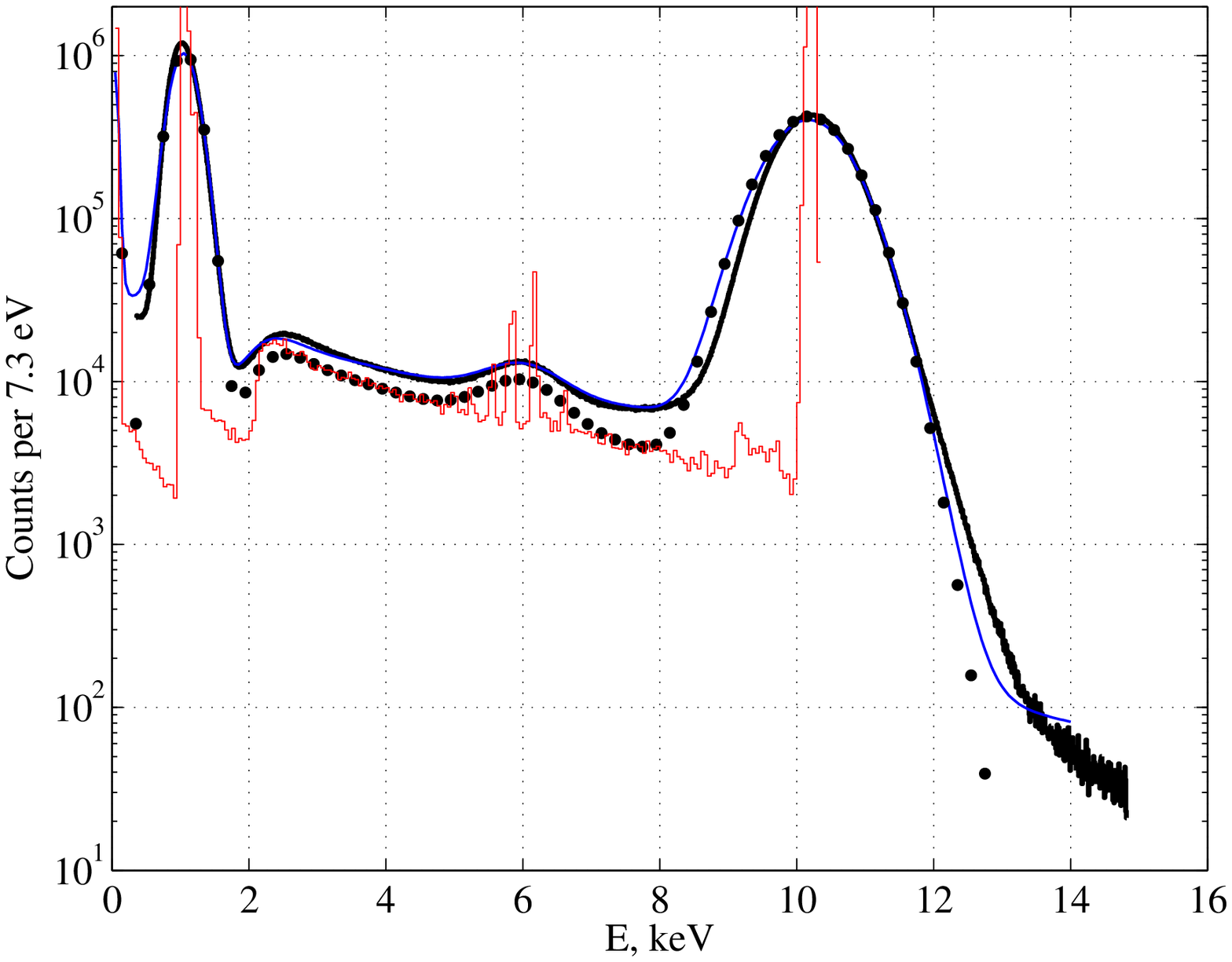}
\caption{\label{fig:model}(Color online) The model response of the counter to decay of $^{37}$Ar (left) and $^{71}$Ge (right) before (upper panel) and after (bottom panel) the patch was applied to the current version of Geant4. The solid black line is the measured spectrum; the jagged thin line (red online) is the model response; and the smooth thin line (blue online) is the model response after distortion. In addition, the model response smeared by the energy resolution $R$ only is shown separately at the bottom panel (black points).}%
\end{figure*}

Another significant source of the response distortion is the end effect resulting from degraded events due to the decrease of the avalanche gain at the counter ends. It can be described in terms of the fraction $D$ of degraded events. The fraction $D$ is normally proportional to the ends' volume, where the electric field becomes nonuniform, divided by the whole counter volume. In this measurement $D$ was about 5\%. A special design of a counter (e.g., with the length increased up to 200~mm) may depress $D$ down to $\sim$1\%. Hereinafter the distribution of degraded events, caused by the end effect, is assumed to have a rectangular shape. In this case, the distortion of the response due to degraded events has the form
\[
S_{\scriptscriptstyle D}(E) = (1-D)S(E) + D {\int_E^{\infty} \frac{S(x) dx}{x}},
\label{eq:degrad}   
\]
where $S_{\scriptscriptstyle D}(E)$ is the response after the distortion.

At high rates pileup can also drastically distort the response. It can be described in terms of the fraction $C$ of indistinguishable coincidences of pulses. The fraction $C$ is usually proportional to the counting rate multiplied by the pulse width. In the case of these measurements, one expects $C$ to be of the order of 0.1\% or less. Analytically this kind of distortion can be expressed as
\[
S_{\scriptscriptstyle C}(E) = (1-C)S(E) + C {\int_0^E S(x) S(E-x) dx},
\label{eq:coinc}   
\]
where $S_{\scriptscriptstyle C}(E)$ is the response after the distortion.

There are another minor sources of uncertainty that can also distort the response. For example, the shape of the tube or anode wire may be slightly nonuniform, but it is not necessary to account for it.

\section{\label{sec:mresult}Results of simulation}

The first simulation immediately revealed nonphysical peaks in the model response produced by Geant4. Later it became clear that the current version of Geant4 does not properly trace the filling of shell vacancies, balancing nevertheless the total energy release with additional local energy deposition. We suggested a code modification that gives a way to solve the problem. The model responses both before and after the patch compared to the measured spectra are presented below. The physical sense of the patch is also described there.

\subsection{\label{subsec:before}Response before patch}

The model response of the counter to $^{37}$Ar and $^{71}$Ge before the patch was applied to the current version of Geant4 is shown in jagged thin line (red online) in Fig~\ref{fig:model} (upper panel). The responses were binned in 50~eV and were distorted as described above with resolution $R$=25\% at 2.8~keV, degraded events $D$=6\%, pileup $C$=0.01\% --- for $^{37}$Ar and $R$=13.5\% at 10.4~keV, $D$=5\%, $C$=0.5\% --- for $^{71}$Ge (smooth thin line, blue online). Note, that the parameters were determined a priori as described above in subsection \ref{subsec:syst}. The exact values of the parameters were obtained from the response after the patch was applied, see subsection \ref{subsec:after}. The counts per energy bin correspond to measured spectra.

In comparison to measured spectra the model responses reveal a strong discrepancy. First, in the original model response just before systematic distortion the K-peak is split onto three main lines. The energy resolution of the counter is sufficient to observe the splitting in the experimental spectrum of $^{71}$Ge if it would exist. In the spectrum of $^{37}$Ar those lines would remain unresolved. Second, in the model response of  $^{71}$Ge the intermediate peak at 2.4~keV, that is observed in the measured spectrum, has not been reproduced. Thus at this stage of simulation it became already clear that the current version of Geant4 somehow proceeds to fill vacancies improperly. Such a conclusion was proved, in particular, by the same splitting that was observed in the primary radiations of $^{37}$Ar and $^{71}$Ge generated by Geant4, just before tracing in the counter (are not shown in figures).

%
%

\subsection{\label{subsec:patch}Geant4 code modification}

A modification was done in the class G4UAto\-mic\-De\-exci\-ta\-tion of the Geant4 that simulates the deexcitation of an atomic shells. When a vacancy appeared in some bottom atomic shell after an electron capture or an escape due to ionization, it cascades to upper shells. If the energy difference is emitted with an X-ray only, the current version of Geant4 will trace this single vacancy moving properly, taking into account all known branching ratios.

In contrast, when an additional vacancy appears due to an Auger electron emission, the Geant4 will not trace its further transitions. The energy imbalance inevitably occuring is compensated by a so-called ``local energy deposition'' being added a posteriori and assumed to be nonionizing.

We have developed a code modification that makes it able to trace the transitions of all possible vacancies.

\subsection{\label{subsec:after}Model response after patch}

The model responses to $^{37}$Ar and $^{71}$Ge after the patch was applied to the Geant4 are represented by jagged thin line (red online) in Fig~\ref{fig:model} (bottom panel). The designations and the systematic distortion are the same as for the upper panel. In addition, the model responses smeared by the energy resolution $R$ only and just before the end effect $D$ and the pileup $C$ distortions are shown separately by black points. We did not perform a full-scale optimization searching for the minimum of chi-squared. Instead we slightly varied the parameters $R$, $D$ and $C$ near the values estimated in the subsection \ref{subsec:syst} together with the intensity of model response in order to fit it to the measured one ``by eye''.

Even without full-scale fit it is clear that after the patch was applied Geant4 reproduces the measured spectra much better. For instance, the degraded events distribution, that has a slope in logarithmic scale in the $^{71}$Ge measured spectrum, is sloped also in the model response. It differs from the distribution of the events degraded at the ends of the counter, that is expected to have a rectangular shape. The effect is due to the near-wall decays when a primary electron releases a part of its initial energy in the wall of counter. In particular, in this case the degraded events are generated by 8~keV KLL Auger electron, being emitted after the K-capture and partially escaping to the wall. If it is gone completely, only the energy of deexcitation of two L-vacancies of the daughter gallium (about 2.4~keV in sum) will remain to be detected. This peak is clearly reproduced in the model response together with the brightly expressed valley between it and the L-peak. The Auger electrons at lower energies, appearing near the wall, reveal the same effect. Such a behavior, with the slope depending on the electron energy, is in strong accordance to the geometry origin of the near-wall degradation.

In smeared model response to $^{71}$Ge, the 6~keV peak is reproduced by the secondary X-ray lines also. This kind of radiation is emitted in the filling of vacancies occuring at the L-shell of xenon ionized by the primary radiations from $^{71}$Ge. The peak is formed by the events where secondary xenon X-rays escape the counter, taking the L-shell binding energy of about 4~keV away. The same peak is also observed in the measured spectrum, and is commonly referred to as the escape peak.

\section{\label{sec:discussion}Discussion}

One may conclude that after the patch is applied, the model response reproduces all visible specific features of the measured spectrum both for $^{37}$Ar and for $^{71}$Ge, even without optimizing the systematic distortion parameters. Not only K- and L-peak with the nearly correct intensity ratios are well reproduced, but also specific features of the $^{71}$Ge spectrum are reflected there, including the near-wall degraded events slope and the intermediate peaks.

The measured spectra of $^{37}$Ar and $^{71}$Ge in a proportional counter thus provide a convincing proof of reliability of the suggested code modification. The proposed patch has been incorporated into Geant4 ver. 10.2, making it able to simulate Auger cascades properly. From the point of view of initial goal of the work one can be sure that the major sources of uncertainty are under almost full control.

In addition, given the widespread use of Geant4, we note that obtained results have a broad range of significance. The correct tracing of deexcitation makes it possible to reproduce the response of any radiation detector at low energy range. This feature is very important in laboratory experiments searching for massive sterile neutrinos as well as for dark matter directly via recoil nuclei (see, for example, \cite{gaitskell:2004dm} and \cite{Agnese:2013jaa}). This work enables physicists to have confidence in Geant4 simulations in the keV energy range.

\section*{Acknowledgments}
We appreciate T.~Tchuvilskaya and the team of the Skobeltsyn Institute of Nuclear Physics (Lomonosov Moscow State University) proton cyclotron for production of $^{37}$Ar and $^{71}$Ge isotopes. We are grateful to V.~Yants for providing us with proportional counters, as well as to V.~Gavrin and other members of the SAGE team for their assistance in low-background measurements. We thank also the Geant4 developer group especially S.~Incerti and L.~Pandola for useful discussions. Finally, we are grateful to J.~Nico for careful reading of the manuscript. This work was supported in part by the Russian Foundation for Basic Research (project No. 14-22-03069).

\bibliography{gnimb}

\end{document}